\documentclass[%
reprint,
showpacs,
 amsmath,amssymb,
aps,
prb,
]{revtex4-1}

\usepackage{amsfonts}
\usepackage{amsmath, amssymb}
\usepackage{bm} 
\usepackage{graphicx} 
\usepackage{color}
\usepackage{ulem}

\newcommand{\veps}{\varepsilon}

\renewcommand{\Im}{\mbox{Im}}
\renewcommand{\Re}{\mbox{Re}}

\begin{document}


\title{Collective polaritonic modes in an array of two-level quantum emitters coupled to optical nanofiber}

\author{D. F. Kornovan$^{1,2}$}
\email{newparadigm.dk@gmail.com}
\author{A.S. Sheremet$^{1,3}$}%
\author{M.I. Petrov$^{1,2,4}$}%
\affiliation{$^{1}$ITMO University, Birzhevaya liniya 14, 199034 St.-Petersburg, Russia}
\affiliation{$^{2}$St.-Petersburg Academic University, 8/3 Khlopina str., 194021 St.-Petersburg, Russia}
\affiliation{$^{3}$Russian Quantum Center, Novaya str. 100, 143025 Skolkovo, Moscow Region, Russia}
\affiliation{$^{4}$University of Eastern Finland, Yliopistokatu 7, FI-80101 Joensuu, Finland} 

\date{\today}

\begin{abstract}
 In this paper we develop a microscopic analysis of the light scattering on a periodic two-level atomic array  coupled to an 	optical nanofiber. We extend the  scattering matrix approach for two-level system interaction with nanofiber fundamental waveguiding mode $HE_{11}$, that allows us modeling the scattering spectra. We support these results  considering the  dispersion of the polaritonic states formed   by the superposition of the fundamental mode of light $HE_{11}$ and the atomic chain states. To illustrate our approach we  start with  considering  a simple  model of light scattering over atomic array in the free space. We discuss the Bragg diffraction at the atomic array and show that the scattering spectrum is defined by the non-symmetric coupling of two-level system with nanofiber and vacuum modes.  The proposed method allows considering two-level systems interaction with full account for  dipole-dipole interaction both via near fields and long-range interaction owing to nanofiber mode coupling.
\end{abstract}

\pacs{78.67.Pt, 03.65.Nk, 42.82.Et}
\maketitle

\section{Introduction}
Controlling interaction of quantum emitters with optical nanostructures at the single-photon level is a key tool for the realization of quantum technologies \cite{Kimble2008, Akimov2007}. {Most experimental efforts focus on the reversible mapping of quantum states between light and matter, and the implementation of quantum networking protocols using this interaction\cite{DeRiedmatten2008,Weber2014}. In this context  localization of photonic modes at the nanoscale {object} opens a feasible route for on-chip quantum communication \cite{Yao2010,Lodahl2015}, and allows implementing  quantum networking protocols \cite{Duan2008,Birnbaum2006}}. At the same time the evanescent character of electromagnetic field {manifested near a nanoobject} reveals fundamentally new features of light-matter interaction \cite{Rivera2016,Bliokh2015}. It is supported  by the recent experimental progress in coupling single quantum sources to surface plasmon polaritons \cite{Kress2015}, and to photonic crystal waveguide modes \cite{Coles2016}, as  well as by the results in  neutral atoms trapping in the vicinity of an optical nanofiber \cite{Kien2004,Nayak2007}. The latter system is a versatile platform for achieving  efficient light-atom coupling due to collective nature of atomic interaction with evanescent field of {the} single photon mode \cite{Balykin2004}. This gives an exceptional opportunity to develop new approaches to study the optical interaction of quantum many body systems at the nanoscale {level}.        


{In this prospective} the interaction between {a} two-level system and the evanescent field of {the}  photonic mode yields to {forming} of mixed polaritonic states with {modified} dispersion relation \cite{Tame2013,Torma2015}. The strong modification of  dispersion is observed in a system of  coupled  plasmonic or dielectric resonators  \cite{Weber2004,Campione2011,Petrov2015, Savelev2015}, which manifest themselves as classically coupled dipole-dipole particles. Nevertheless,  considering the cold atomic system trapped in the vicinity of an optical nanofiber the origin of the  polaritonic states and their dispersion is significantly overlooked. The existing theoretical approaches base on reflection and transmission spectroscopy of an incident fiber mode \cite{LeKien2014a}. The {theoretical} predictions \cite{LeKien2005,Russell2009},  and experimental verification \cite{Nayak2007} have shown that spectral distribution of atomic fluorescence is {strongly} affected by the presence of the nanofiber. It has been experimentally examined \cite{Kluge2016} by detecting  the Bragg diffraction in the atomic chain. Despite its universality and technical convenience this approach does not clarifies the exact picture of atom-atom interaction in the presence of a nanofiber as omits the exact details of dipole-dipole coupling. This paper is aimed on eliminating this gap by considering the eigenstates of the atomic array coupled to the nanofiber modes, which manifest themselves as polaritonic states. We apply the $T$-matrix method for studying the scattering of the nanofiber mode over the constructed polaritonic states. In contrary to reflectance and transmittance spectroscopy approach this method can be universally extended to an arbitrary dense atomic array. In order to expose the full picture of the atom-photon interaction, we start our consideration with single photon scattering at the atomic chain in vacuum and identifying the polaritonic states.

 The manuscript is organized as follows: is Sec.~\ref{Approach} we describe in details theoretical approach of the considered problem in the case of the atomic chain in the vacuum and in the vicinity of the nanofiber; in Sec.~\ref{vacuum} we discuss the calculated scattering cross sections and interpret them using polaritonic band diagram; Sec.~\ref{fiber} we extend the approach for the case of  {the} nanofiber and observe strong backscattering into the nanofiber mode when the Bragg condition is satisfied.
\begin{figure}[htbp]
\includegraphics[width=0.9\columnwidth]{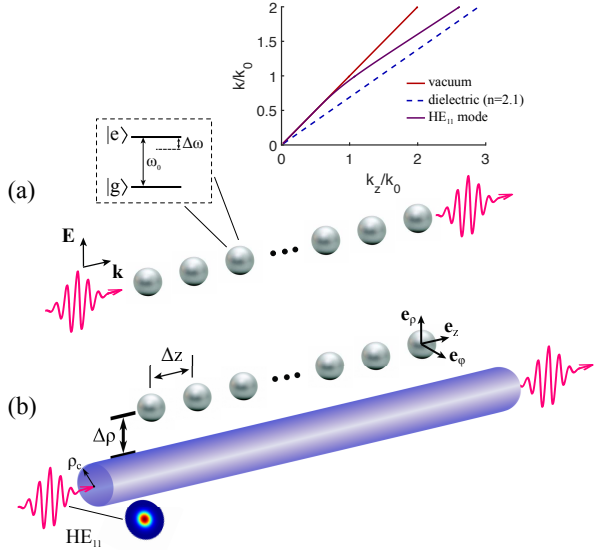}
\caption{  Light scattering on the $1D$ array of two-level atoms with the period $\Delta z$: (a)  A single photon with polarization vector being parallel to the dipole moment of the atomic transition {scatters and propagates} along the atomic chain axis	. (b)   {The scattering of a quasi-circularily polazired single  photon from the fundamental guided mode $HE_{11}$ on the array of atoms trapped in the vicinity of the optical nanofiber. All atoms are positioned at the same distance $\Delta\rho$ from the fiber surface}.}
\label{Scheme}
\end{figure}

\section{Theoretical  approach}
\label{Approach}

We consider the light scattering on an one-dimensional (1D) array  of $N$ two-level atoms with a   {period}  $\Delta z$   {and compare this process for two systems:}  i)  {the} atomic chain  in the vacuum{,} see Fig. \ref{Scheme} (a), ii) and in the vicinity of an optical silica  nanofiber  {($n=2.1$)}{,} see Fig. \ref{Scheme} (b). In  {the first} case  we consider a single photon  {scattering} with  {a} wave vector   {directing} along the  {atomic} chain, and in the presence of  {a} nanofiber we consider   {the propagation of a guided light field } in the fundamental  {mode} $HE_{11}${,} Fig. \ref{Scheme} (b).  All  atoms are placed at the same distance $\Delta\rho=0.3\lambda_0$ from the fiber surface with  radius $\rho_{c}=0.25\lambda_0$, which is a typical value for such systems realized in the experiment \cite{Goban2012}. Here   $\lambda_0$ is the wavelength of the atomic transition. 


\subsection{  {Interaction of a single photon with an atomic chain in the vacuum}} 
\label{ApproachVac}
 
  {In the microscopic quantum theory}  {light} scattering process  {can be described using the standart} $T$-matrix  {formalism} \cite{ClaudeCohen-Tannoudji2004}.   {The total} Hamiltonian $\hat H$  {describing the interaction between propagating light and the atomic chain can be expanded in}  a sum of   {the non}-perturbed part $\hat H_{0}$ and  {the} interaction term $\hat V$  {such that $\hat{H}=\hat{H}_{0}+\hat{V}$, where}:
\begin{eqnarray}
\hat H_{0} &=& \sum_{n}\hbar \omega_0 \hat \sigma^+_n \hat \sigma^-_n + \sum_{\mu}\hbar \omega_{k} \hat a^\dagger_{\mu}\hat a^{ }_{\mu}\nonumber,\\
\hat V &=&  -\sum_{n} \mathbf{\hat d}_{n} \mathbf{\hat E} (\mathbf{r}_n).
\label{Hamiltonian1}
\end{eqnarray}
Here  the interaction part of the Hamiltonian $\hat V$  {is considered} in the dipole approximation, where $\mathbf{\hat d}_{  {n}}$ is the transition dipole moment operator of the $n$-th atom,  {$\hat{\sigma}^{+}_{n} = |e_{n}\rangle \langle g_{n}|$ and $\hat{\sigma}^{-}_{n} = |g_{n}\rangle \langle e_{n}|$ are raising and lowering atomic operators,}  {$\hat{a}^{\dagger}_{\mu}(\hat{a}_{\mu})$ are the bosonic creation (annihilation) operators, index}  $\mu$  describes a particular field mode $\mu = \left(\mathbf{k}, s \right)$, where $\mathbf{k}$ is the wavevector, $s=1,2$ denotes two orthogonal polarizations and $\mathbf{\hat E} (\mathbf{r}_n)$ is the total microscopic electric field operator  {which can be written as}
\begin{eqnarray}
\hat {\mathbf{E}} (\mathbf{r}) = \sum_{\mu}\sqrt[]{\frac{2\pi \hbar \omega_{k}}{\mathbb{V}}}\left( i\mathbf{e}_\mu \hat a_{\mu}e^{i\mathbf{kr}} + h.c.\right),
\end{eqnarray}
  {where $\mathbb{V}$} is the quantization volume,  and $\mathbf{e}_\mu$ is the unit polarization vector.

The $T$-matrix then can be written in the form\cite{ClaudeCohen-Tannoudji2004}: 
\begin{eqnarray}
\label{Tmatrix}
\hat{T} = \hat{V} + \hat{V} \hat{G}(E + i0) \hat{V},
\end{eqnarray}
where $\hat G( z ) = (z - \hat{H})^{-1}$ is the resolvent operator of the total Hamiltonian.  {In accordance with the rotating wave approximation} the matrix elements of the $\hat{T}$-operator  {can be found as a projection onto the Hilbert subspace of the vacuum state for the field subsystem and the single excited state for the atomic subsystem:
\begin{equation}
\hat{P} \hat{G}(E) \hat{P} = \hat{P}\; \dfrac{1}{E - \hat{H}_0 - \hat{\Sigma}(E)}\; \hat{P},
\label{Projection}
\end{equation}
where we defined the projector operator as following:
\begin{equation}
\hat P = \sum_{n=1}^N |g_1, ..., e_n, ... ,g_N; \{0_{\mu}\}\rangle \langle \{0_{\mu}\}; g_1, ..., e_n, ..., g_N|.
\end{equation}
}

  {In the Eq. (\ref{Projection}) we introduced the level-shift operator} $\hat{\Sigma}$  \cite{ClaudeCohen-Tannoudji2004}. The form of  {this} operator can be found as perturbative series in powers of $\hat{V}$. 

In the lowest order of the perturbation theory  {the operator} $\hat{\Sigma}$   {can be described by} two contributions  {corresponding to the single-particle and the double-particle interactions} \cite{Sheremet2012}. The single particle contribution leads to the Lamb shift and the finite lifetime of the  {atomic} excited state, while the double{-}particle contribution is responsible for the excitation transfer between the atoms.



Here we work in the resonant approximation, which allows considering the scattering of a photon  {with a carrier frequency $\omega$ }  close to the atomic transition  {frequancy $\omega_0$}. In this approximation  $\hat{\Sigma} (E)$ is assumed as a slowly varying function of the argument so that $\hat{\Sigma}(E)\approx \hat{\Sigma}(E_0)$. The single and double particle contributions   {can be written as}:
\begin{eqnarray}
\Sigma^{(nn)}(E_0) &=& \hbar\left(\Delta_L - i \dfrac{\gamma_0}{2}\right) 
\nonumber\\
\Sigma^{(mn)}(E_0) &=& - \mathbf{d}^{*}_{m}\bigg[ \frac{e^{ikR}}{R}\bigg(\left(1+\dfrac{ikR-1}{k^2R^2}\right)\mathbf{I}+ 
\nonumber\\ 
&&\dfrac{\mathbf{R}\otimes\mathbf{R}}{R^2}\cdot \dfrac{3-3ikR-k^2R^2}{k^2R^2}\bigg)\bigg]\mathbf{d}_n,
\end{eqnarray}
where $\Delta_L$ is the Lamb shift, $\gamma_0$ is the spontaneous emission rate, $k=\omega/c$ is the wavenumber of a vacuum photon, $R=|\mathbf{r_m}-\mathbf{r_n}|$ is the distance between atoms $m$ and $n$, $\mathbf{I}$ is the unit dyad, and $\otimes$ stands for the outer product.

Once the operator matrix $\hat \Sigma$  is computed we can construct the denominator in \eqref{Projection} and, by inverting it, obtain the matrix for projected resolvent and the $T$-matrix. We are interested in the scattering of the  photon back into the same field mode, which is an elastic scattering channel, corresponding to the diagonal matrix element of the $T$-matrix $T_{ii}(E)$. This matrix element is connected with the total scattering cross section according to the optical theorem \cite{Sheremet2012, Ezhova2016}: $\sigma_{tot}(E) \sim \Im \: T_{ii}(E)$. We introduce the normalized total cross section in a following way
 \begin{eqnarray}
\sigma_{N}(E) = \frac{\Im \; T_{ii}^{(N)}(E)}{\Im \; T^{(1)}_{ii}(E_{res})},
\label{sigman}
\end{eqnarray}
where  $\Im T^{(1)}_{ii}(E_{res})$ corresponds to the maximal scattering cross section of a single photon on a single atom.

\subsection{ {Interaction of a guided light with an atomic chain in the presence of a nanofiber}}
However, to correctly  {take} into account  the optical fiber we need to  modify the approach discussed in Sec.~\ref{ApproachVac} and we will do it in two steps. First, we need  modify the ``outer'' operators $\hat V$  in Eq. (\ref{Tmatrix}), which are responsible for the absorption of the incoming  {guided} photon and emission of the photon back into the same field mode. To describe the field subsystem at this step we use the quantization scheme proposed in \cite{Minogin2010}, where the quantized  electric field of the guided mode of the nanofiber can be written as:
\begin{eqnarray}
\mathbf{\hat E} (\mathbf{r}) = \sum_{\mu} \mathbf{E}_\mu(\mathbf{r}) \hat a_{\mu} + h.c.,
\end{eqnarray}
where $\mathbf{E}_\mu$ is the electric field of the guided mode $\mu$:
\begin{eqnarray}
\mathbf{E}_\mu (\mathbf{r}) = i \; \sqrt[]{\frac{2\pi \hbar \omega_\mu}{ \mathbb{L}}} \tilde{\mathbf E}_\mu(\rho, \phi) e^{if\beta_\mu z+im\phi}.
\end{eqnarray}
Here $\beta_\mu$ is the propagation constant, $\tilde{\mathbf E}_\mu(\rho, \phi)$ is the amplitude of the electric field, $\mathbb{L}$ is the quantization length, $f$ and $m$ define the direction of propagation and the mode angular momentum, correspondingly. The electric field is periodic in  $z$-direction and the periodicity condition can be written as $\beta_l \mathbb{L}=2\pi l$, where $l$ is a positive integer number.
The electric field amplitude is normalized according to
\begin{eqnarray}
\int\limits_{0}^{2\pi}\int\limits_{0}^{\infty}\ |\tilde{\mathbf E}_\mu(\rho, \phi)|^2 d\phi \rho d\rho = 1.
\end{eqnarray}

At the next step, we need to calculate the matrix elements of  {the operator} $\hat{\Sigma}$ in the presence of  {a} nanofiber. To account for  the excitation transfer between the atoms through the radiation of vacuum modes and modes of  {the} nanofiber we need { to introduce} the proper quantum-electrodynamical description of the electromagnetic field, which was developed by D.-G. Welsch   {\textit{et al.} in} Ref. \onlinecite{Gruner1996}.   {Using this formalism we can modify the Hamiltonian (\ref{Hamiltonian1}) to describe our system as follows}: 
\begin{eqnarray}
\hat{H}_0 &=& \sum\limits_n \hbar \omega_0 \hat{\sigma}^{+}_n \hat{\sigma}^{-}_n + \int d\mathbf{r^{\prime}} \int\limits_0^{\infty} d\omega^{\prime} \hbar \omega^{\prime} \hat{\mathbf{f}}^\dagger (\mathbf{r^{\prime}},\omega^{\prime}) \hat{\mathbf{f}} (\mathbf{r^{\prime}},\omega^{\prime}),
\nonumber \\ 
\hat{V} &=& -\sum\limits_n \hat {\mathbf d}_{n} \hat {\mathbf E}(\mathbf{r}_n),
\end{eqnarray}
where $\omega_0$ is the atomic transition frequency, $ \hat{\mathbf E}(\mathbf{r}_n)$ is the total electric field, $\hat{\mathbf{f}} (\mathbf{r^{\prime}}, \omega^{\prime}), \hat{\mathbf{f}}^{\dagger} (\mathbf{r^{\prime}}, \omega^{\prime})$ are the bosonic vector local-field operators, which obey the following commutation relations:
\begin{eqnarray}
\left[\hat{f}_i(\mathbf{r^{\prime}}, \omega^{\prime}), \hat{f}^{\dagger}_k(\mathbf{r}, \omega)\right] &=& \delta_{ik}\delta(\mathbf{r^{\prime}} - \mathbf{r})\delta(\omega^{\prime} - \omega),
\nonumber\\
\left[\hat {f}_i(\mathbf{r^{\prime}}, \omega^{\prime}), \hat{f}_k(\mathbf{r}, \omega) \right] &=& 0
\end{eqnarray}
Positive frequency part of the total electric field has the following form:
\begin{multline}
\hat{\mathbf{E}}^{+}(\mathbf{r}) = i \; \sqrt[]{4 \hbar} \int d\mathbf{r^{\prime}} \int\limits_{0}^{\infty} d\omega^{\prime}\frac{{\omega^{\prime}}^2}{c^2} \sqrt{\veps_I (\mathbf{r^{\prime}}, \omega^{\prime})} \times \\ 
\mathbf{G}(\mathbf{r}, \mathbf{r^{\prime}}, \omega^{\prime}) \hat{\mathbf{f}}(\mathbf{r^{\prime}}, \omega^{\prime}),
\end{multline}
 {where} $\veps_I(\mathbf{r}^{\prime}, \omega^{\prime})$ is the imaginary part of the dielectric permittivity of the media, $\mathbf{G}(\mathbf{r}, \mathbf{r}^{\prime}, \omega^{\prime})$ is the classical Green's tensor of the electric field. In the presence of  {the} optical fiber the Green's tensor can be expanded into 
\begin{eqnarray}
\label{Greens}
\mathbf{G}(\mathbf{r}, \mathbf{r}^{\prime}, \omega)  = \mathbf{G}_0(\mathbf{r}, \mathbf{r}^{\prime}, \omega) + \mathbf{G}_s(\mathbf{r}, \mathbf{r}^{\prime}, \omega),
\end{eqnarray}
where $\mathbf{G}_0$ is the vacuum Green's tensor, and $\mathbf{G}_s$ is the Green's tensor corresponding  to the light scattering from the fiber. The scattering term of the Green's tensor can be expanded into the Vector Wave Functions (WVF) and the details of these calculations are given in the Appendix~\ref{AppendixGreens}.
\begin{figure}[htbp]
\begin{center}
\includegraphics[scale=0.5]{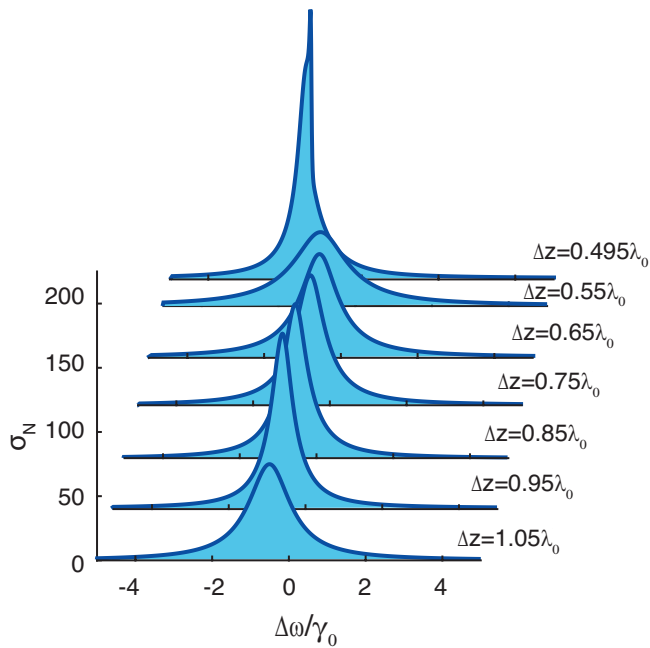}
\caption{Normalized total scattering cross sections   {dependence on photon frequency} detuning  {$\Delta\omega = \omega - \omega_{0}$} for different periods  of the chain  {$\Delta z$}. The dipole transition is oriented parallel to the field polarization ${\bf d}||{\bf E}$. The number of atoms  {is} equal to $N=100$. }
\label{CrSecVac}
\end{center}
\end{figure}
In the lowest non-vanishing order the matrix elements of the level-shift operator in this case can be  {written} as
\begin{eqnarray}
\langle f| \hat \Sigma (E) |i\rangle =\sum_{|\alpha \rangle , |\beta \rangle} \langle f|\hat V|\alpha \rangle \langle \alpha | \frac{1}{E - \hat H_0 + i\eta}|\beta \rangle \langle \beta |\hat V|i\rangle,  \nonumber \\
 \end{eqnarray}
where $|i\rangle$ and $|f\rangle$ are the initial and final states of the system correspondingly,  $|\alpha \rangle, |\beta \rangle$ are the  {two possible intermediate} states with a single elementary excitation for the field subsystem and both atoms are either in the excited or the ground state:
\begin{eqnarray}
|e_{n},e_{m}\rangle \times \hat {\mathbf{f}}^\dagger (\mathbf{r^\prime},\omega^\prime)|\{0\}\rangle, \nonumber \\
|g_{n},g_{m}\rangle \times \hat {\mathbf{f}}^\dagger (\mathbf{r^\prime},\omega^\prime)|\{0\}\rangle.
\end{eqnarray}
The derivation of these matrix elements of the level-shift operator can be found elsewhere \cite{Dung2002,Marocico2009} and here we provide only the final expression:
\begin{eqnarray}
\label{SigmaOp}
\langle f| \hat \Sigma(E) |i\rangle = - 4\pi \frac{\omega_0^2}{c^2} \mathbf{d}^{*}_{m} \mathbf{ G}(\mathbf{r}_m,\mathbf{r}_n,\omega_0)\mathbf{d}_{n}.
\end{eqnarray}
 
   We should notice that in the case of single particle contribution, when $|i\rangle = |f\rangle$ and, thus, $\mathbf{r}_n=\mathbf{r}_m$ the homogeneous part of the Green's function has a singularity in the real part $\Re \left[\mathbf{G}_0(\mathbf{r}_n,\mathbf{r}_n,\omega_0) \right] \rightarrow \infty$ which corresponds to the infinite Lamb shift due to the interaction with the vacuum modes. This term is renormalized and can be thought of as it is already incorporated into the definition of the transition frequency of atomic dipoles $\omega_0$. However, $\Re \left[\mathbf{G}_s(\mathbf{r}_n,\mathbf{r}_n,\omega_0) \right]$ is finite and it leads to a presence of the Lamb shift due to the interaction of the excited atom with the fiber modes. 

Now using  (\ref{SigmaOp}) we can find the  {matrix} $\Sigma(E)$, the $T$-matrix elements and, consequently, the scattering cross section $\sigma_{N}(E)$. In this case when calculating the denominator of Eq. (\ref{sigman}) the atom is placed at the same distance $\Delta\rho$ from the fiber surface as atoms in our periodic chain. Also, we notice that $E_{res}$ now differs from $\hbar \omega_0$  because of the Lamb shift.

\section{Results: Atomic chain in the vacuum}
\label{vacuum}
We consider  the photon scattering   {on} the atomic chain in  {the} vacuum in the geometry shown in Fig.~\ref{Scheme} (a). In this case we assume that the dipole moments of  {the} atoms are aligned parallel to  photon polarization. 

We have applied the $T$-matrix approach to plot the spectra of scattering cross section for different  interatomic distances. The  scattering intensity is shown in Fig.~\ref{CrSecVac}. One can notice that it changes in a non-monotonous way as distance between the atoms varies. The most pronounced changes are observed when the period is approximately $m{\lambda_0}/{2}$, where $m$ is an integer. For instance, changing the   {interatomic} distance from $\Delta z =0.49\lambda_0$ to  $\Delta z =0.55\lambda_0$  results in  decreasing of the intensity and widening of the peak. Similar effect but much weaker is observed when changing the distance from $0.95\lambda_{0}$ to $1.05\lambda_{0}$. This behavior is related to opening of the diffraction channels each times when the Bragg condition is satisfied. On the other hand, this  process can be easily understood by analyzing the  eigenstates of the atomic system, which   manifest themselves in polaritonic states.

\subsection{Polaritonic states in  {the} atomic chain }

The polaritonic states can be constructed by defining the eigenstates of the level-shift operator, which is in our approximation the operator of dipole-dipole atomic coupling. In  the limit of resonant excitation the eigenproblem can be formulated as follows: 
\begin{eqnarray}
\label{EigEq}
\Sigma(\omega_{0}) {\bf v}= \mathcal{E} {\bf v}.
\end{eqnarray}
Here $\Sigma(\omega_{0})$ is the  matrix representation of the level-shift operator. The solution of this equation gives us $N$ complex eigenvalues $\mathcal{E}_i=\hbar \omega_{i}$ and column eigenvectors ${\bf v}_i$, which are the energies and eigenstates of the system described in the basis of states with a single atomic excitation. We utilize the solution of a finite eigensystem to plot the dispersion curve for the infinite chain  \cite{Weber2004}. For that we correlate the eigenvector with the corresponding wavenumber $k_{z}$ by the enumerating the eigenstates in accordance with the  number of nodes $l$ in the profile of the eigenmode ${\bf v_i}$. Then we can assign the corresponding wavenumber $k_z$ to each mode according to: 

\begin{eqnarray}
\dfrac{k_z}K =\frac{(N-2)(l+1)+1}{2N(N-1)},
\end{eqnarray}
where $K=2\pi/\Delta z$ is the reciprocal lattice vector of a periodic chain, and $l=0,1,2..$ is the mode number.   
\begin{figure}[h]
\begin{center}
\includegraphics[scale=0.5]{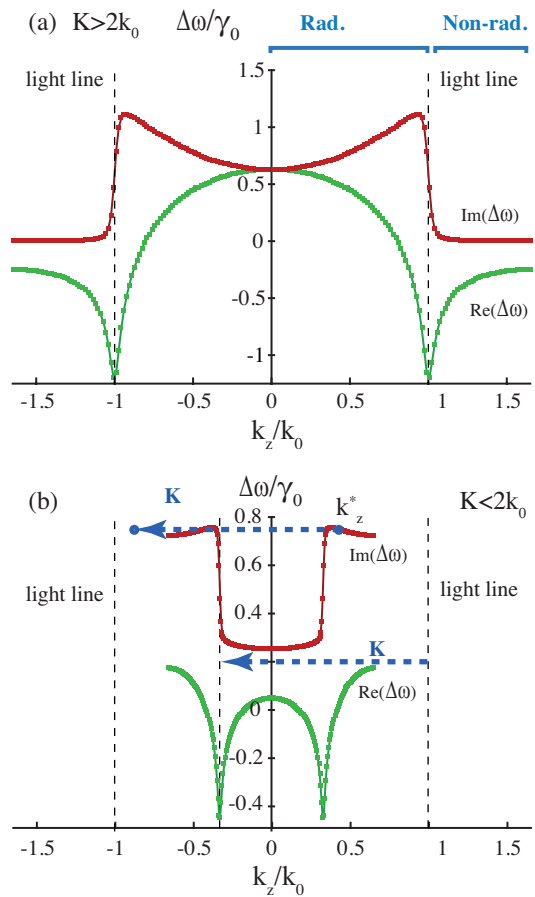}
\caption{The real (red) and imaginary (green) parts of  {the} eigenfrequencies  of the transversal polaritonic states with $\mathbf{d} \perp \mathbf{e_z}$ versus the corresponding $k_z$ values for: (a) sub-diffractional case ${K>2k_{0}}$ ($\Delta z=0.3 \lambda_{0}$); (b) diffractional case ${K<2k_{0}}$ ($\Delta z=0.75 \lambda_{0}$).  The dispersion of  {the} vacuum photon modes (light line) are shown with dashed lines. The regions of radiative and non-radiative states are marked. }
\label{Disp_vac_T}
\end{center}
\end{figure}
This procedure allows us  plotting both real and imaginary parts of  {the} eigenfrequencies of our system as functions of $k_z$, where the real part accounts for the dispersion of normal modes and the imaginary part describes radiative losses or the inverse lifetimes of the eigenstates.

\begin{figure}[htbp]
\begin{center}
\includegraphics[scale=0.5]{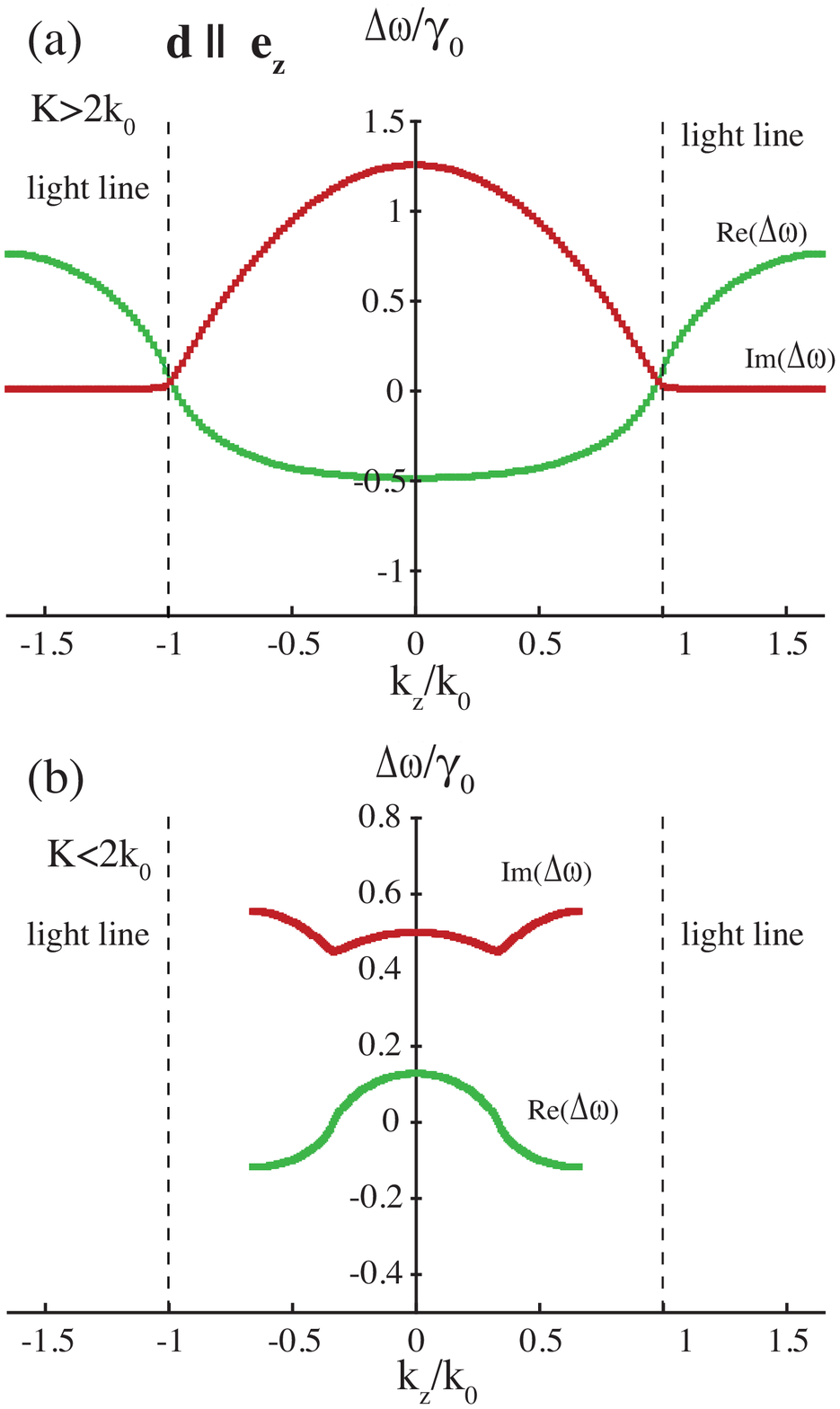}
\caption{The real (red) and imaginary (green) parts of  {the} eigenfrequencies of the  {longitudial} polaritonic states with $\mathbf{d} \parallel \mathbf{e_z}$ versus the corresponding $k_z$ values for: (a) sub-diffractional case ${K>2k_{0}}$ ($\Delta z=0.3 \lambda_{0}$); (b) diffractional case ${K<2k_{0}}$ ($\Delta z=0.75 \lambda_{0}$). The dispersion of  {the} vacuum photon modes (light line) are shown with dashed lines.}
\label{Disp_vac_L}
\end{center}
\end{figure}

In order to support the scattering cross section spectra shown in~Fig.~\ref{CrSecVac} we  illustrate the light interaction with the atomic chain by  plotting the dispersion curves for transversal ($\mathbf{d} \perp \mathbf{e_z}$) polaritonic states{,} see Fig.~\ref{Disp_vac_T}.  We consider  (a) sub-diffractional (${K>2k_{0}}$)  and (b) diffractional (${K<2k_{0}}$) regimes, when the first Bragg condition is satisfied. The light line, which is vertical on a scale of the polaritonic bandwidth as $\gamma_0 \ll \omega_0$, divides the states into radiative and non-radiative ones. In the vicinity of point  {$k_{z}=k_{0}$} the atomic states  undergo hybridization with vacuum photon modes. For the diffractional case{, see} Fig.~\ref{Disp_vac_T} (b), all the eigenmodes become radiative as they appear above the light line. However, one should notice that hybridization features do  preserve, but are shifted from the light line for quantity $K$, as the wave vector  $k_{z}$ is a quasi-vector of polaritonic state and is conserved up to a reciprocal vector.  Moreover, the states near the band edges ($k_{z}>K-k_{0}$)  become more radiative than states in the band center as they have two channels of radiation: (i) they can emit a photon with  $k_{z}^{ph}=k_z^{*}$  and (ii)  a photon with $k_z^{ph}=k_{z}^{*}-K$. 

\begin{figure}[t]
\begin{center}
\includegraphics[scale=0.4]{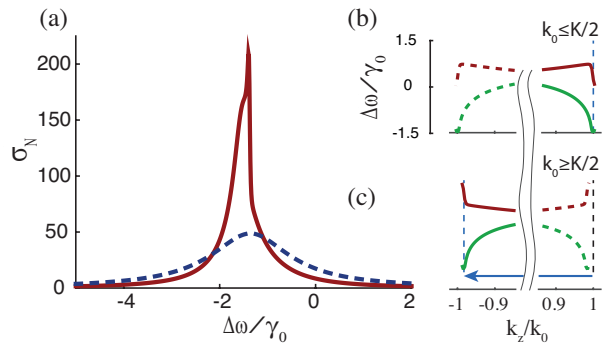}
\caption{The illustration of the  diffraction channel opening in the photon scattering  on two-level atomic array in vacuum.  (a) The scattering cross sections for two chain periods: $K\gtrsim 2k_{0}\ (\Delta z=0.49\lambda_0)$ (red solid line) and $K\lesssim 2k_{0}\ (\Delta z=0.51\lambda_0)$ (blue dashed line). b) and c) show dispersion curves (red) and inverse lifetimes (green) of these states in the region close to the  situations shown in (a).  The number of atoms is $N = 100$.}
\label{CS_vac_switch}
\end{center}
\end{figure}

The dispersion of  {the} longitudinal modes ($\mathbf{d} \parallel \mathbf{e_z}$), similarly to the transversal modes, can be also divided into radiative and non-radiative regions, see Fig.~\ref{Disp_vac_L}. However,  the hybridization with  {the} vacuum modes is in the vicinity of light line is not observed due to polarization mismatch: the vacuum modes have traversal polarization and the polaritonic excitations are longitudinal.
 
\subsection{Bragg diffraction}

The  {plotted} dispersion curves clarify the character of the  cross-section spectra shown in Fig.~\ref{CrSecVac}, in particular, the  opening of the first  Bragg  diffraction channel when the period changes from $\Delta z=0.49 \lambda_{0}$ to $\Delta z=0.51 \lambda_{0}$.  The  $z$-component of the incident photon equals to  $k_{0}$ according to Fig.~\ref{Scheme} (a) and for sub-diffractional regime the scattering occurs on the states at the edge of the band{,} see Fig.~\ref{CS_vac_switch} (b). In the sub-diffractional regime, when $k_{0}\lesssim K/2$ these states have low losses, that  provides narrow cross section spectrum shape{,} see solid line in Fig.~\ref{CS_vac_switch} (a). After switching to diffractional regime $k_{0}\gtrsim K/2$ the incident photon scatters off the states with $k_{z}=k_{0}-K$  (umklapp process) as shown in Fig.~\ref{CS_vac_switch} (c). Due to high radiative losses connected to free space diffraction  the cross section spectra is wide, see dashed line in Fig.~\ref{CS_vac_switch} (a).


\section{Results: Atomic chain in the vicinity of the optical nanofiber}
\label{fiber}
The presence of  {an} optical  {nano}fiber changes the character of atomic interaction and allows long-range dipole-dipole coupling between the atoms not only via  {the} vacuum, but also through the waveguiding mode. To study this effect and its influence on the scattering of the waveguiding mode over atomic chain, we have applied the $T$-matrix method. In contrast to commonly used  {the} transfer matrix method, where  {the} interaction of  {the} guiding mode with each atom is treated individually \cite{Kien2015,Kluge2016}, here we consider the scattering on collective polaritonic states with account for  {the} full atomic dipole-dipole interaction and splitting their energy levels. For that we start with building the eigenstate picture of the atomic system with  {the} optical  {nano}fiber.

\subsection{Dispersion of polaritonic states}
\begin{figure}[h]
\includegraphics[width=0.4\textwidth]{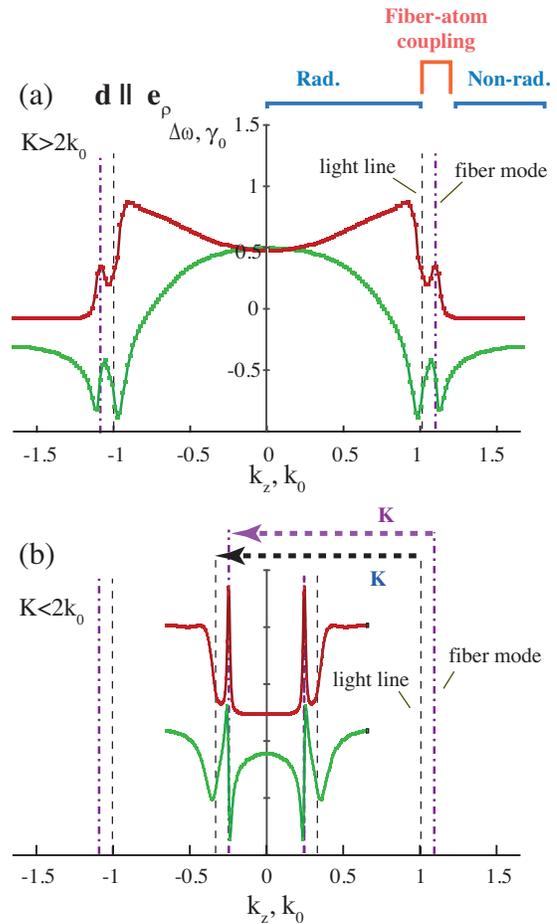}
\caption{The real (red) and imaginary (green) parts of the eigenfrequencies  of the transversal polaritonic states with $\mathbf{d} \parallel \mathbf{e_\rho}$ versus the corresponding $k_z$ values for: (a) sub-diffractional case ${K>2k_{0}}$ ($\Delta z=0.3 \lambda_{0}$); (b) diffractional case ${K<2k_{0}}$ ($\Delta z=0.75 \lambda_{0}$).  The dispersion of vacuum photon modes (light line) are shown with black dashed lines. The dispersion of the nanofiber fundamental mode $HE_{11}$ is shown with purple dash-dot line. The regions of radiative, non-radiative, and strongly  coupled to the nanofiber mode states are marked.   {The number of atoms is $N = 100$, the nanofiber radius is $\rho_c = \lambda_0/4$, the distance from the fiber surface is $\Delta\rho = 0.3\lambda_0$.} }
\label{DispFiber}
\end{figure}

The polaritonic dispersion relation in the presence of  {the} optical  {nano}fiber can be found from  {the} eigenstates of the system (\ref{EigEq}), but with the corrected level-shift operator, which includes interaction with the  {nano}fiber by means of scattering Green's function in Eq. (\ref{Greens}). The real and imaginary parts of eigenfrequencies versus the corresponding  $k_{z}$-values are plotted in Fig.~\ref{DispFiber} for  transverse $\mathbf{d} \parallel \mathbf{e_{\rho}}$  modes. The parameters of the  {nano}fiber are chosen in the way that it supports only one fundamental mode $HE_{11}$ at the frequency of atomic transition $\omega_{0}$. The fiber mode dispersion curve is shown with the dash-dot line in Fig.~\ref{DispFiber} in addition to the vacuum photon line shown with the dashed line. In the sub-diffractional regime $K>2k_{0}$ the  {nano}fiber interaction channel gives anti-crossing-like feature in the polaritonic dispersion in the vicinity of  $k_{z}=k_{0}^{f}$, where $k_0^f$ denotes the wavevector of the waveguiding photon having frequency $\omega_0$.  The nanofiber modifies the non-radiative atomic states and forms  {nano}fiber coupled polaritonic states{,} see Fig.~\ref{DispFiber} (a). These states are situated closely to radiative states as wavevector of the fundamental waveguiding mode is close to the wavevector of the vacuum photon $|k_{0}-k_{0}^{f}|\ll k_{0}$ (see Fig.~\ref{Scheme}). The peak in the  spectrum of  {the} imaginary frequency at $k_{z}=k_{0}^{f}$ is related to the leakage of the state through the fiber mode.  For the diffractional regime $K<2k_{0}${,} see Fig.~\ref{DispFiber} (b){,} all states become radiative and  there is resonant anti-crossing coupling to the waveguiding mode of the fiber at $\pm k_{0}^{f}\mp K$ along with the vacuum mode coupling at $\pm k_{0}\mp K$. 

\begin{figure}[htbp]
\includegraphics[width=0.4\textwidth]{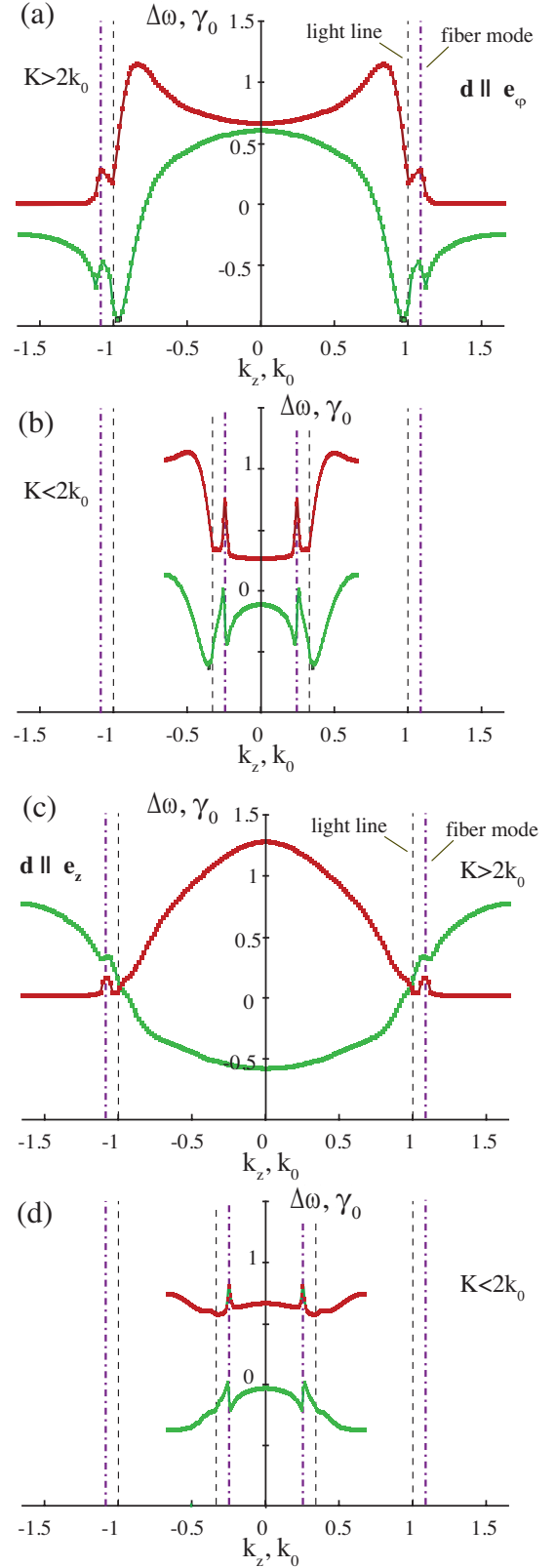}
\caption{The real (red) and imaginary (green) parts of  {the} eigenfrequencies  of the transversal polaritonic states with $\mathbf{d} \parallel \mathbf{e_\varphi}$ (a, b) and  {the} longitudinal states with $\mathbf{d} \parallel \mathbf{e_z}$ (c, d) versus the corresponding $k_z$ values are shown. (a, c) sub-diffractional case ${K>2k_{0}}$ ($\Delta z = 0.3 \lambda_{0}$); (b, d) diffractional case ${K < 2k_{0}}$ ($\Delta z = 0.75 \lambda_{0}$).  The parameters and notation are the same as in Fig.\ref{DispFiber}. }
\label{DispFiber2}
\end{figure}

The field of the  {fundamental}  fiber mode  {$HE_{11}$} has all three components of the electric field, thus, in  {the} general all of them contribute to the dipole-dipole interaction. For the completeness of the consideration we have plotted the rest two polarizations  of  the dipole moments of the atomic transition: azimuthal transversal ($\mathbf{d} \parallel \mathbf{e_{\varphi}}$)  and longitudinal    ($\mathbf{d} \parallel \mathbf{e_{z}}$) are shown in Fig.~\ref{DispFiber2}. The dispersion of azimuthal modes  is similar to radial modes but has weaker interaction with  {the} fiber mode due to the weaker amplitude of the azimuthal component of electrical field in the fiber mode. The longitudinal modes fully resembles the longitudinal modes in vacuum with  {the} fiber mode interaction being weaker than for  {the} transversal modes. However,  there is no coupling of atoms with vacuum field due to polarization mismatch, but atoms are interacting with the fiber mode, see Fig.~\ref{DispFiber2}(c,d), as $HE_{11}$ mode is not fully transversal and has nonzero $z$-component of the electric field, which gives its contribution to the interaction constant.

\subsection{ {The }fiber mode scattering}

\begin{figure}[htbp]
\begin{center}
\includegraphics[scale=0.45]{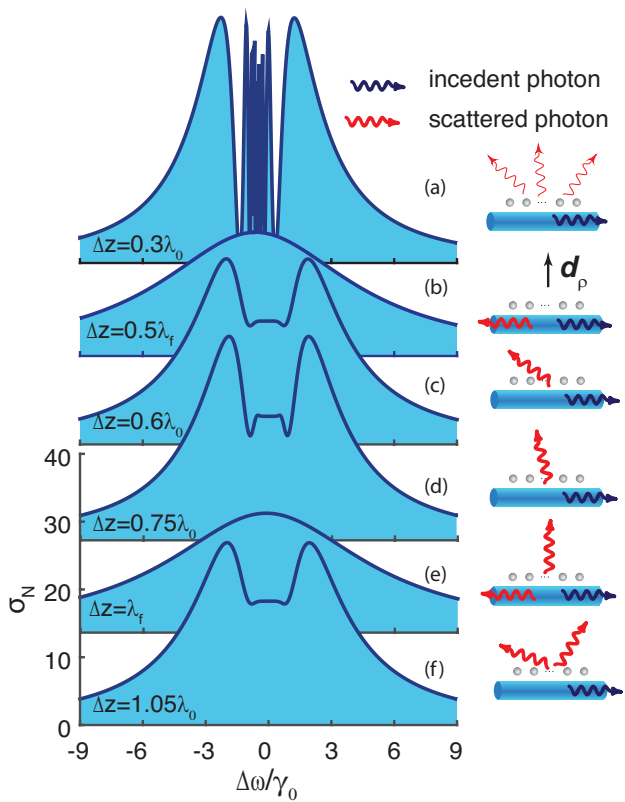}
\caption{  {The} normalized scattering cross section spectra of two-level atomic chain consisting of $N=200$ atoms   {in the vicinity of} the nanofiber for different periods $\Delta z$: a) $0.3 \lambda_{0}$, b) $0.5 \lambda_{f}$, c) $0.6 \lambda_{0}$, d) $0.75 \lambda_{0}$, e) $\lambda_{f}$, f) $1.05 \lambda_{0}$. The nanofiber radius is $\rho_c = \lambda_0/4$, the distance from the fiber surface is $\Delta\rho = 0.3\lambda_0$. }
\label{CS_fiber}
\end{center}
\end{figure}

We have analyzed the scattering of the  {fundamental} fiber mode $HE_{11}$ by the atomic chain in sub-diffractional and diffractional regimes as shown in Fig.~\ref{CS_fiber}. We consider all atoms having only ${\bf d_{\rho}}$ component of dipole transition matrix elements, which corresponds to Fig.\ref{DispFiber}. The presence of the nanofiber makes the system effectively one dimensional, which leads to significant changes in scattering cross section spectra if compared to the vacuum case. We plot the scattering cross section spectrum, which corresponds to probability of a single photon to escape from the waveguiding mode after  interaction with the atomic chain.

One can see from Fig.~\ref{CS_fiber} that for  {the} sub-diffractional regime the spectrum is modulated by sharp resonances in the vicinity of atomic resonant frequency $\omega_{0}$. These resonances correspond to scattering on the states with $k_{z}\approx k_{f}$ having low losses. Though these states are below the light line so they have finite radiational lifetime due to the finite length of the chain{,} see Fig.~\ref{DispFiber} (a). 
 When the Bragg condition {$\Delta z=0.5\lambda_{f}$} is satisfied the spectrum becomes purely {Lorentzian, that is defined by the existing highly radiative state of the atomic system, and the main channel is back scattering into the waveguiding mode, propagating in the direction opposite to the incident.}   
The scattering process for $K<2k_{0}$ goes through the umklapp process as shown in Fig.~\ref{DispFiber} (b) with dashed purple arrow, and corresponds to a vacuum diffraction with a specific $k_{z}$. The scattering spectra acquires constant region in its central part with the oscillatory features at the edges as shown in Fig.~\ref{CS_fiber}. The further increase of the chain period results in almost periodic change of the cross section spectra, and, in particular, when $\Delta z=\lambda_{f}\ (K=k_{f})$ we have  {the} Bragg condition of the second order and backscattering into the waveguide mode with $k_{z}=-k_{f}$.
   
\section{Discussion}

The cross section spectra plotted in Fig.~\ref{CS_fiber} have  two qualitatively distinct  profiles: (i) the Lorentzian shape profile if the condition of  {the} fiber Bragg diffraction is satisfied and{,} see Fig.~\ref{CS_fiber}(b,e); (ii)  {a} profile with notch in the middle of the spectrum{, see} Fig.~\ref{CS_fiber}(a,c,d,f). The  Bragg diffraction is associated with the scattering on highly radiative state which appears at the edge of the band similar to the case shown in Fig.~\ref{CS_vac_switch}. The incident photon is scattered by radially oriented dipole moment back into the waveguiding mode of the nanofiber. However, for the other periods the photon is diffracted  in the cone with a fixed angle, defined by the condition $k_{z}=k_{f}-K$ as it is schematically shown in the right column of Fig.~\ref{CS_fiber}.      

\begin{figure}[htbp]
\begin{center}
\includegraphics[scale=0.47]{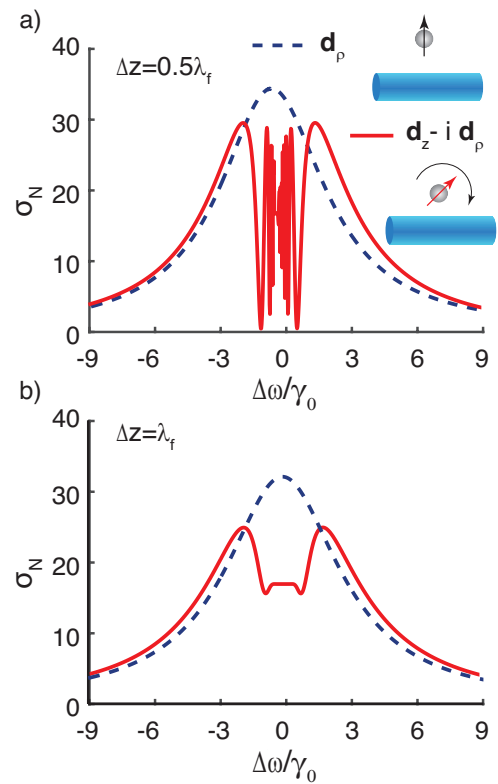}
\caption{ {The} normalized scattering cross section spectra of  {the} atomic chain consisting of $N=200$ atoms with radially polarized  (dashed line) and $\sigma_{+}$ polarized dipole moments (red solid line) in the regime of  a) first ($\Delta z = 0.5\lambda_{f}$) and b) second ($\Delta z = \lambda_{f}$) fiber Bragg diffraction. The parameters are the same as in Fig.\ref{CS_fiber}.}
\label{CS_fiber_asym}
\end{center}
\end{figure}

The change of the spectrum shape we associate with the switching of the diffraction from symmetric (in  {the} case of diffraction into  {the} fiber mode) into the asymmetric scattering (diffraction into the vacuum modes). The asymmetry in photon emission by an excited atom in the vicinity of a nanofiber has been actively discussed recently \cite{Junge2013,Mitsch2014,Kien2015}. In particular, it was shown \cite{Kien2015} that an atom with transversal and longitudinal components of the dipole moment  has asymmetry in forward and backward spontaneous emission rate into  {the} nanofiber mode. This results in the asymmetry of the single atom reflectance of  {the} waveguding mode propagating in  {the} forward or backward directions also known as spin-locking effect \cite{Bliokh2015}.  Accounting on that in  {the} case of asymmetric emission rate the Bragg reflection is suppressed and notched reflectance spectra \cite{Kien2015} is observed. The asymmetry in the case shown in Fig.~\ref{CS_fiber} (a, c, d, f) can be explained by the  asymmetry in the emission rate  of  {the} collective polaritonic states into  the vacuum and  fiber mode. When the scattering goes back into the fiber mode exactly at the Bragg resonance the symmetry is conserved, but at the vacuum diffraction this symmetry breaks. To support this statement we compared scattering of the incident photon on atoms with (i) radial component of the dipole moment  ${\bf d_{0}}$ and (ii) $\sigma_{+}$ polarized dipole having both radial ${\bf d_{\rho}}$ and $z$-component ${\bf d}_{z}$.  In the latter case the two components have $\pi/2$ phase shift but the absolute dipole moment equals $|{\bf d_{\rho}}-i{\bf d}_{z}|/\sqrt{2}=d_{0}$. Contrary to radially polarized atom the $\sigma_{+}$  atoms has  {the} strong asymmetry in coupling with the forward and backward propagating fiber mode \cite{VanMechelen2016}. We have made the calculation for the fiber backscattering regime, when the symmetry should be conserved for linearly polarize atom, but not for the circularly polarized atom. In Fig.\ref{CS_fiber_asym} the scattering cross section is shown for the case of  {the} first and second fiber Bragg resonance in  {the} case of radially polarized atoms (dashed line) and in  the case of $\sigma_{+}$ atoms. We see the pronounced switching from  {the} Lorentzian spectral shape to a notched shape. For the first fiber Bragg condition $\Delta z=0.5\lambda_{f}$, see Fig.\ref{CS_fiber_asym} (a), one can see sharp peaks in the center of the band due to scattering by  {the} long-lived collective atomic states. In the case of  {the} second Bragg resonance{,} see Fig.\ref{CS_fiber_asym} (b) there the sharp peaks are smeared out as all the polaritonic state are above light cone and, thus, have high losses.

\section{Conclusions} 

In this work we have considered a single photon scattering on {an} ordered finite chain of two-level atoms embedded in  {the} vacuum or  {trapped} in the vicinity of  {the} single mode dielectric nanofiber. We have developed the scattering matrix technique  and analyzed the scattering cross section spectrum of a single photon in the presence of the nanofiber. This approach allowed  us incorporating the atomic dipole-dipole interactions both via vacuum near fields and long range coupling through the waveguiding mode. To support the results of our simulations we have constructed the polaritonic states of the interacting  atomic array both in  vacuum and close to the nanofiber, that has not been done before for the considered type of quantum system. The obtained dispersion curves for the polaritonic states allowed us interpreting the results of the scattering cross section calculations and demonstrated the effects of atomic array coupling with single waveguiding mode of the nanofiber. Finally, we have shown that the photon scattering over the atomic chain in  the presence of the	 nanofiber is influenced by the effects of spin-locking coupling of atoms with nanofiber  and vacuum modes. The proposed approach, which combines constructing the polaritonic eigenstates of the atomic system with the quantum scattering theory, can be effectively applied for modeling experiments in the light interaction with quantum systems at the nanoscale level. 

\section*{Acknowledgements}

The work was supported by Russian Fund of Basic Research within the project 16-32-60167. M.I.P. acknowledges support from Academy of Finland, grant No. 288591.  A.S.S. acknowledges ITMO Fellowship and Visiting Professorship Program for financial support.

\appendix
\section{Green's tensor}
\label{AppendixGreens}

 In order to obtain the $\Sigma^{mn}$ matrix elements we need to construct the Green's tensor of the system, which can be found from:  

\begin{eqnarray}
\left[- \frac{\omega^2}{c^2}\veps(\mathbf{r},\omega) + \mathbf{\nabla} \times \mathbf{\nabla} \times  \right] \mathbf{G}(\mathbf{r},\mathbf{r}^\prime,\omega)=\mathbf{I}\delta(\mathbf{r}-\mathbf{r}^\prime),
\label{A1}
\end{eqnarray}
where $\veps(\mathbf{r},\omega)$ is the complex dielectric function and $\mathbf{I}$ is the unit dyad. In our case we consider a dielectic cylindrical waveguide of radius $\rho_c$ and dielectric permittivity $\veps$ being constant inside the cylinder. To find the solution we apply the scattering superposition method \cite{Marocico2009,Marocico2011}, which allows us expanding the Green's tensor into the homogeneous and inhomogeneous terms:
\begin{equation}
\mathbf{G}(\mathbf{r},\mathbf{r}^\prime,\omega) = \mathbf{G}_0(\mathbf{r},\mathbf{r}^\prime,\omega) + \mathbf{G}_s(\mathbf{r},\mathbf{r}^\prime,\omega).
\end{equation}

As soon as we consider atomic dipoles in the vicinity of the waveguide, so that $\mathbf{r}, \mathbf{r}^\prime$ are outside the cylinder, the homogeneous term is always present and describes the field generated directly by the source placed at the point $\mathbf{r}^\prime$ at the field point $\mathbf{r}$. This term can be obtained analytically from the Green tensor written in cartesian coordinates using the transformation from cartesian to cylindrical coordinates $\mathbf{S}(\phi)\mathbf{G}_{0}^{Cart}(\mathbf{r},\mathbf{r}^\prime,\omega)\mathbf{S}^T(\phi)$, where $\mathbf{G}_{0}^{Cart}$ has an analytic expression \cite{Novotny2012} and is given by 
\begin{eqnarray}
\mathbf{G}_{0}^{Cart}(\mathbf{r}, \mathbf{r^\prime}, \omega) = \left( \mathbf{I} + \frac{1}{k^2}\mathbf{\nabla} \otimes \mathbf{\nabla}\right)G_0(\mathbf{r},\mathbf{r}^\prime,\omega),
\end{eqnarray}
here $G_0(\mathbf{r},\mathbf{r}^\prime,\omega)$ is the  Green's function of the scalar Helmholtz equation.

The scattering term can be calculated via the integral representation of the homogeneous part. To obtain this representation we apply the method of VWF explained in details in Ref.~\onlinecite{Chew1999,Tai1994}, here we cover only the basic ideas and provide the final expressions.
 To find the solution of the vector Helmholtz equation \eqref{A1} we introduce the scalar Helmholts equation and  the solution of this equation in the cylindrical coordinates:
 \begin{eqnarray}
&& \nabla^2\phi(\mathbf{k},\mathbf{r}) + k^2\phi(\mathbf{k},\mathbf{r}) =0, 
\nonumber\\
&& \phi_n(k_{z},\mathbf{r})=J_n(k_\rho \rho)e^{in\theta+ik_zz},
 \end{eqnarray}
here $J_n(x)$ is the Bessel function of the first kind, $\mathbf{r} = (\rho,\theta,z)$ are the cylindrical coordinates and $k_\rho$, $k_z$ are the projections of the wavevector $\mathbf{k}$.
The solution of the vector Helmholtz equation may be written in terms of the following vector wavefunctions:
\begin{eqnarray}
\mathbf{M}_n(k_z,\mathbf{r}) &=& \mathbf{\nabla} \times [\phi_n(k_z,\mathbf{r})\mathbf{e_z}] 
\nonumber\\
\mathbf{N}_n(k_z,\mathbf{r}) &=& 
\frac{1}{k}\mathbf{\nabla } \times \mathbf{M}_n(k_z,\mathbf{r}) 
\end{eqnarray}
where $\mathbf{e_z}$ is the so-called pilot vector, the unit vector pointing in the $z$ direction. These WVFs $\mathbf{M}$, $\mathbf{N}$ correspond to $TE/TM$ modes of the field.

One can show  \cite{Chew1999} that the homogeneous part of the Green's function can be expanded in terms of these vector wavefunction in the following way:
\begin{multline}
\mathbf{G}_{h}(\mathbf{r},\mathbf{r^\prime}, \omega)=-\dfrac{\mathbf{e_{\rho}e_{\rho}}}{k_0^2} \delta (\mathbf{r}-\mathbf{r^\prime}) + \\
+\dfrac{i}{8\pi}\sum_{n= - \infty}^{\infty} \int\limits_{-\infty}^{\infty} \frac{dk_z}{k_{0\rho}^2} \mathbf{F}_n (k_z, \mathbf{r}, \mathbf{r^\prime})
\end{multline}
and the $\mathbf{F}_n (k_z, \mathbf{r}, \mathbf{s})$ function is given by
\begin{eqnarray}
\begin{cases}
 \mathbf{M}_n^{(1)} (k_z, \mathbf{r})\overline{\mathbf{M}}_n (k_z, \mathbf{r^\prime}) + \mathbf{N}_n^{(1)} (k_z, \mathbf{r})\overline{\mathbf{N}}_n (k_z, \mathbf{r^\prime})& \nonumber \\
 \mathbf{M}_n (k_z, \mathbf{r})\overline{\mathbf{M}}_n^{(1)} (k_z, \mathbf{r^\prime}) + \mathbf{N}_n (k_z, \mathbf{r})\overline{\mathbf{N}}_n^{(1)} (k_z, \mathbf{r^\prime})& 
\end{cases}
\end{eqnarray}

here the first line holds for $\rho_r>\rho_{r^\prime}$ while the second one for $\rho_r<\rho_{r^\prime}$, and $k_0={\omega}/{c}$, $k_{0\rho }= \sqrt{k_0^2-k_z^2}$ and the superscript $(1)$ in vector wave functions denotes that the Bessel function of the first kind $J_n(k_\rho \rho)$ should be replaced with the Hankel function of the first kind $H^{(1)}_n(k_\rho \rho)$. Here we provide the explicit form of WVF:

\begin{eqnarray}
 \mathbf{M}_n(k_z,\mathbf{r}) &=& 
\begin{pmatrix}
\frac{in}{\rho} J_n(k_{0\rho } \rho)\\
- k_{0\rho } (J_n(k_{0\rho } \rho))'\\
0
\end{pmatrix} e^{i n \theta + i k_z z},
\nonumber\\ 
\mathbf{N}_n(k_z,\mathbf{r}) &=& 
\begin{pmatrix}
\frac{ik_z k_{0\rho}}{k} (J_n(k_{0\rho} \rho))'\\
-\frac{n k_z}{\rho k} J_n (k_{0\rho } \rho)\\
\frac {k_{0\rho}^2}{k} J_n (k_{0\rho} \rho)
\end{pmatrix} e^{i n \theta + i k_z z} 
\nonumber\\
\overline{\mathbf{M}}_n(k_z,\mathbf{r^\prime}) &=& 
\begin{pmatrix}
-\frac{in}{\rho^\prime} J_n(k_{0\rho } \rho^\prime)\\
- k_{0\rho } (J_n(k_{0\rho } \rho^\prime))'\\
0
\end{pmatrix}^T e^{- i n \theta^\prime - i k_z z^\prime},
\nonumber\\ 
\overline{\mathbf{N}}_n(k_z,\mathbf{r^\prime}) &=& 
\begin{pmatrix}
-\frac{ik_z k_{0\rho }}{k} (J_n(k_{0\rho } \rho^\prime))'\\
-\frac{n k_z}{\rho^\prime k} J_n (k_{0\rho } \rho^\prime)\\
\frac {k_{0\rho }^2}{k} J_n (k_{0\rho } \rho^\prime)
\end{pmatrix}^T e^{- i n \theta^\prime - i k_z z^\prime}
\notag
\end{eqnarray}
where $J_n(k_\rho \rho)'$ means derivative with respect to the dimensionless argument.

Now having the integral representation of the homogeneous term of the Green's function, we can construct the scattering term in a similar fashion. Let us denote the medium outside the dielectric cylinder as $1$ and the medium inside as $2$. The particular form of the Green's tensor depends on the position of a source point $\mathbf{r^\prime }$: whether it is inside or outside the cylinder. As soon as we are interested in a situation, when both source and receiver are outside the cylinder and in the latter we consider only the second case. Thus, the total Green's tensor can written as:
\begin{eqnarray}
\begin{cases}
\mathbf{G}^{11}(\mathbf{r},\mathbf{r^\prime},\omega) = \mathbf{G}^{11}_h(\mathbf{r},\mathbf{r^\prime},\omega) + \mathbf{G}^{11}_s(\mathbf{r},\mathbf{r^\prime},\omega), 
\nonumber\\
\mathbf{G}^{21}(\mathbf{r},\mathbf{r^\prime},\omega) = \mathbf{G}^{21}_s(\mathbf{r},\mathbf{r^\prime},\omega), 
\nonumber
\end{cases}
\notag
\end{eqnarray}
here two superscripts denote position of the receiver the source point respectively and the two scattering parts of the Green's tensor has the following form:
\begin{eqnarray}
\mathbf{G}_{s}^{11}(\mathbf{r,r^\prime,\omega}) &=&
\frac{i}{8 \pi} \sum_{n= - \infty}^{\infty} \int\limits_{-\infty}^{\infty} \frac{dk_z}{k_{\rho1}^2} \mathbf{F}^{11 (1)}_{\mathbf{M};n, 1}(k_z,\mathbf{r})\overline{\mathbf{M}}_{n,1}^{(1)}(k_z,\mathbf{r^\prime}) 
\nonumber \\  
&+&\mathbf{F}^{11 (1)}_{\mathbf{N};n,1}(k_z,\mathbf{r})\overline{\mathbf{N}}_{n,1}^{(1)}(k_z,\mathbf{r^\prime}) ,
\nonumber\\
\mathbf{F}^{11 (1)}_{\mathbf{M};n,1}(k_z,\mathbf{r}) &=& R^{11}_{MM}\mathbf{M} ^{(1)}_{n,1}( k_z, \mathbf{r})+R^{11}_{NM} \mathbf{N}_{n,1}^{(1)} (k_z,\mathbf{r}) ,
\nonumber\\
\mathbf{F}^{11 (1)}_{\mathbf{N};n,1}(k_z,\mathbf{r}) &=& R^{11}_{MN}\mathbf{M} ^{(1)}_{n,1}(k_z, \mathbf{r})+R^{11}_{NN} \mathbf{N}_{n,1}^{(1)} (k_z,\mathbf{r}).
\nonumber\\
\mathbf{G}_{s}^{21}(\mathbf{r,r^\prime,\omega}) &=&
\frac{i}{8 \pi} \sum_{n= - \infty}^{\infty} \int\limits_{-\infty}^{\infty}  \frac{dk_z}{k_{\rho1}^2} \mathbf{F}^{21}_{\mathbf{M};n,2}(k_z,\mathbf{r})\overline{\mathbf{M}}_{n,1}^{(1)}(k_z,\mathbf{r^\prime}) 
\nonumber \\ 
&+& \mathbf{F}^{21}_{\mathbf{N};n,1}(k_z,\mathbf{r})\overline{\mathbf{N}}_{n,1}^{(1)}(k_z,\mathbf{r^\prime}) ,
\nonumber\\
\mathbf{F}^{21}_{\mathbf{M};n,2}(k_z,\mathbf{r}) &=& R^{21}_{MM}\mathbf{M}_{n,2}(k_z, \mathbf{r})+R^{21}_{NM} \mathbf{N}_{n,2} (k_z,\mathbf{r}) ,
\nonumber\\
\mathbf{F}^{21}_{\mathbf{N};n,2}(k_z,\mathbf{r}) &=& R^{21}_{MN}\mathbf{M}_{n,2}(k_z, \mathbf{r})+R^{21}_{NN} \mathbf{N}_{n,2} (k_z,\mathbf{r}),
\nonumber
\end{eqnarray}
here the scattering Fresnel coefficients $R_{AB}^{ij}$ are introduced and the second subscript in the VWFs denotes that $k$ and $k_\rho$ should be replaced with their values inside the corresponding media $k_i=\veps_i(\mathbf{r},\omega)k_0$, $k_{\rho i}=\sqrt[]{k_i^2 - k_z^2}$. We should notice that unlike the case of homogeneous term, here we have products of $\mathbf{M}$ and $\mathbf{N}$, which is due to the fact that the normal modes in our case have hybrid nature.

The form of the Fresnel coefficients mentioned above can be found by imposing the boundary conditions on the Green's tensor at the surface of the cylinder 
\begin{eqnarray}
{\begin{cases}
\mathbf{e}_{\rho} \times [\mathbf{G}^{11}(\mathbf{r},\mathbf{r^\prime},\omega) - \mathbf{G}^{21}(\mathbf{r},\mathbf{r^\prime},\omega) ]|_{\rho_r = \rho_c} = 0, 
\nonumber \\
\mathbf{e}_{\rho} \times \mathbf{ \nabla_r } \times [\mathbf{G}^{11}(\mathbf{r},\mathbf{r^\prime},\omega) - \mathbf{G}^{21}(\mathbf{r},\mathbf{r^\prime},\omega)]|_{\rho_r = \rho_c} = 0
\end{cases}}
\end{eqnarray}

Solving for this, we can find the Fresnel coefficients $R_{AB}^{ij}$ and, finally, construct the scattering part of the Green's tensor $\mathbf{G}_s(\mathbf{r},\mathbf{r}^\prime,\omega) $. 
\bibliographystyle{apsrev4-1}
\bibliography{Rfrncs2}

\end{document}